\documentstyle[11pt,newpasp,twoside,psfig,epsf]{article}
\def\edcomment#1{\iffalse\marginpar{\raggedright\sl#1\/}\else\relax\fi}
\reversemarginpar

\begin{document}

\title{The Heating of the ICM: Energy Crisis and viable solutions} 
\author{Paolo Tozzi}
\affil{Oss. Astronomico di Trieste, via G.B. Tiepolo 11, 34100
Trieste -- Italy}

\begin{abstract}
X--ray observations indicate that non--gravitational processes play a
key role in the thermodynamics of the Intra Cluster Medium (ICM).  The
effect of non--gravitational processes is imprinted in the ICM as an
entropy minimum, whose effects are visible in the
Luminosity--Temperature relation and in the Entropy--Temperature
relation.  However, the X--ray emission alone cannot discriminate
between different mechanisms and sources of heating.  There are no
answers at present to the following questions: how much
non--gravitational energy per baryons is present in the ICM?  When was
this energy injected?  Which are the sources of heating?  The
embarrassment in front of these questions is amplified by the fact
that the most viable sources of heating, SNae and stellar winds, seem
to be inefficient in bringing the ICM to the observed entropy level.
We may call it the {\sl energy crisis}.  Here we review the main
aspects of this crisis, listing possible solutions, including
other sources, like AGNs and Radio Galaxies, or other mechanisms, like
large scale shocks and selective cooling.
\end{abstract}

\section{Evidences for non--gravitational heating}

Observations in the X--ray band provided convincing evidences for
non--gravitational heating of the diffuse baryons in the potential
wells of groups and clusters of galaxies (ICM).  Among them: the shape
of the L--T relation (steeper than the self--similar behaviour
$L\propto T^2$ predicted in the case of gravitational processes only)
and the entropy excess in the center of groups, recently found by
Ponman, Cannon \& Navarro (1999).  Non--gravitational heating of the
ICM is expected also on the basis of observations of an average
metallicity $Z\simeq 0.3 ~ Z_\odot$.  In fact, SNae and stellar winds
are the most viable source of heating and metal enrichment of the
ICM. It should be understood how much of the ejected energy goes into
the ICM and if it is enough to generate the observed entropy plateau.

\section{How the entropy works}

If the excess entropy is present in the baryons before collapse ({\sl
external heating}, or preheating, see Tozzi \& Norman 2001, hereafter
TN), it will be preserved in the cores of dark matter halos after
virialization.  In virtue of the extra pressure support, in fact, the
gas is accreted adiabatically without shock heating.  The excess
entropy also decreases the density in the central regions.  This, in
turn, rapidly decreases the X--ray luminosity which is proportional to
the square of the density.  The effect is stronger in small groups,
where the energy responsible for the entropy is comparable to the
gravitational one, while clusters, where gravity is dominant, are
mostly unaffected.  This produces a steepening of the $L$--$T$
relation and breaks the scale invariance in the density profiles of
clusters and groups.  The entropy has also the effect of suppressing
the radiative cooling in the central regions.  This model allows to
trace the evolution of X--ray luminosity and temperature of groups and
clusters after assuming a Press \& Schechter--like law for the
accretion rate of baryons (see TN).  In Figure 1 we show that the
evolution corresponds to tracks moving along the local $L$--$T$
relation.  Thus, a constant $L$--$T$ is predicted up to $z\simeq 1$,
in agreement with observations (Mushotzky \& Scharf 1997, Borgani et
al. 2001).  The external entropy level that satisfies the observations
is in the range $K_* = 0.2 - 0.3 \times 10^{34}$ erg cm$^2$ g$^{-5/3}$
(see TN).

\begin{figure}
\plottwo{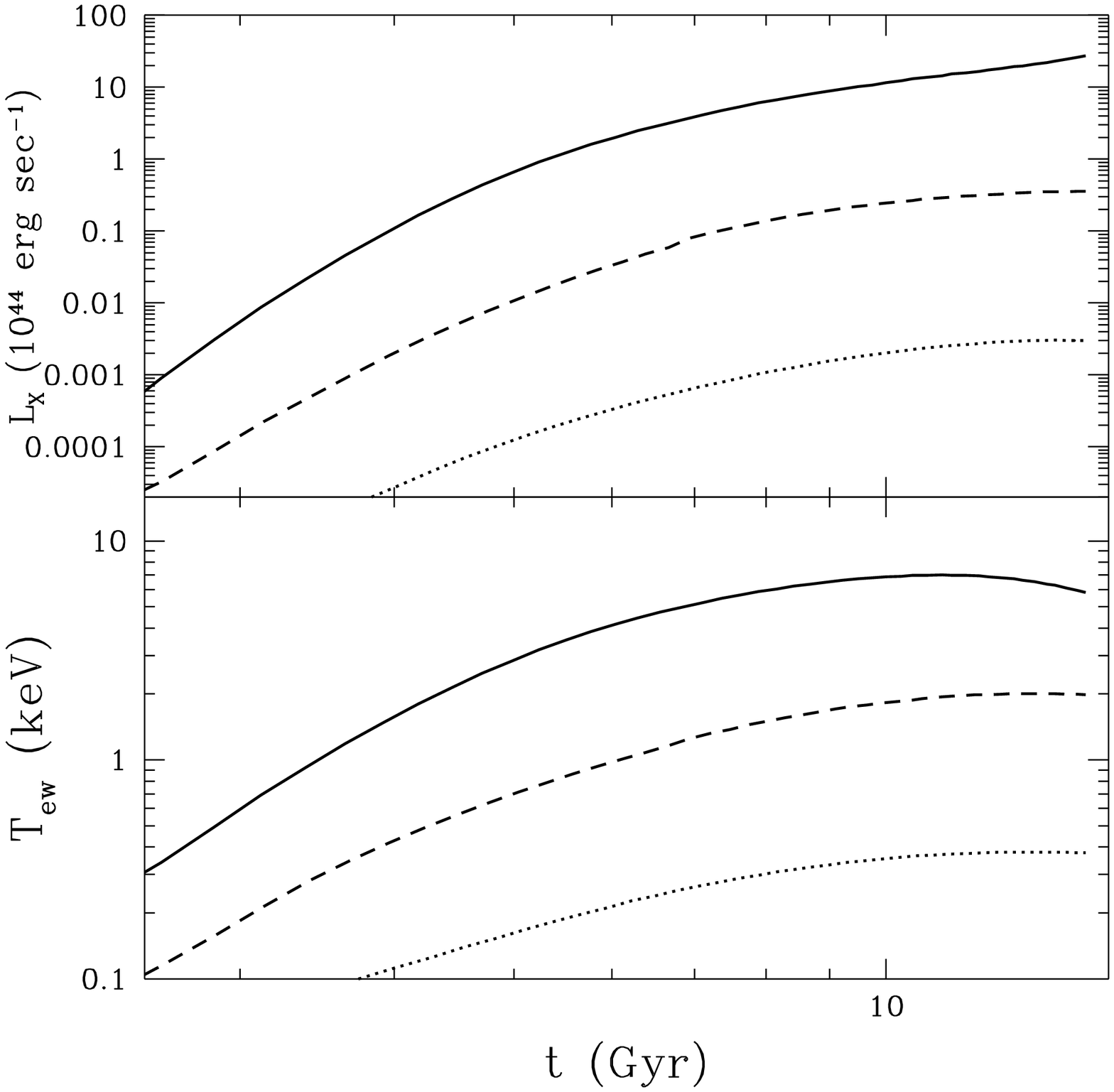}{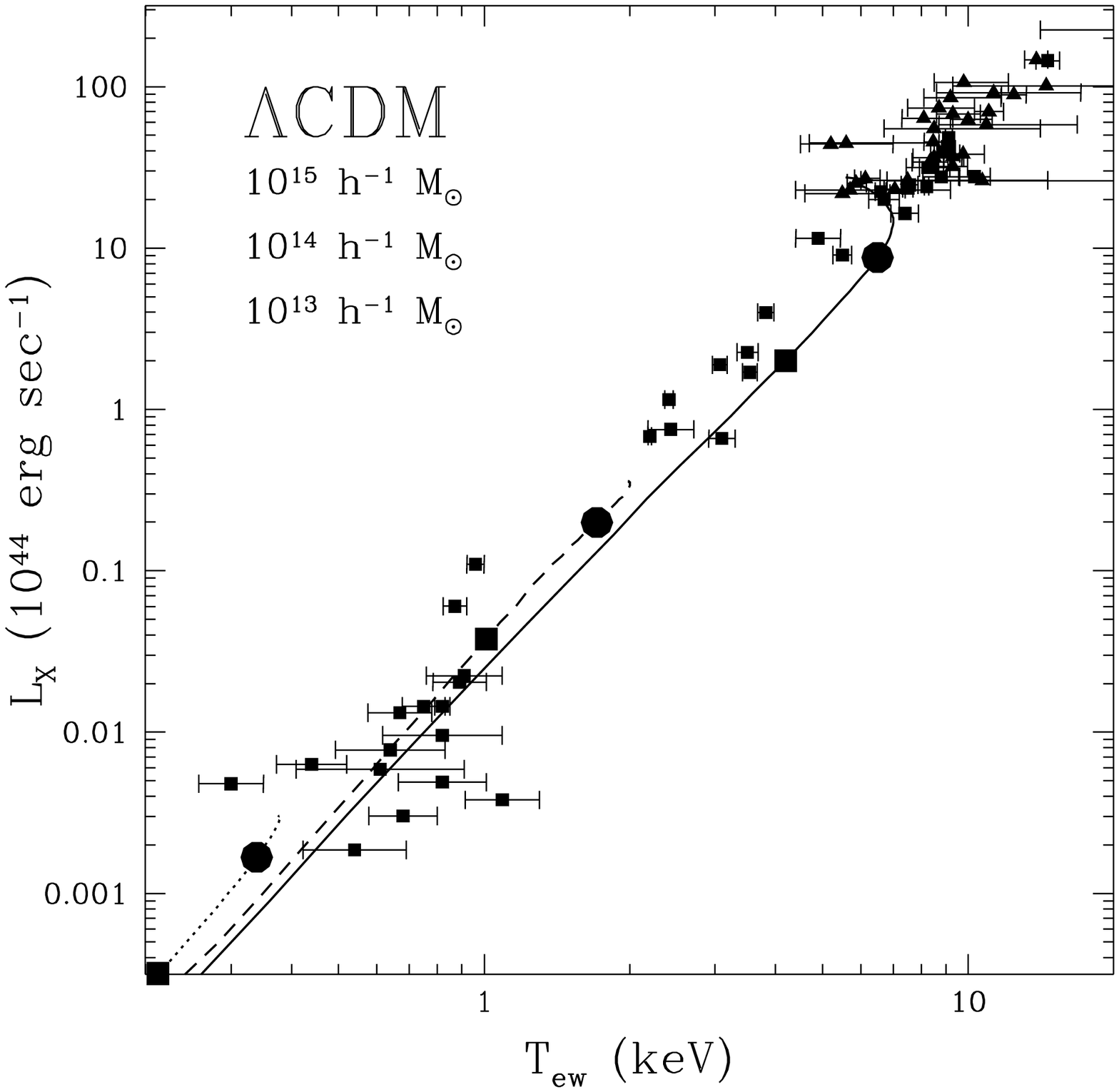}
\caption{Left: The evolution of the bolometric luminosity $L_X$ and of
the emission--weighted temperature $T_{ew}$ is shown as a function of
time for a final mass of $10^{15} - 10^{14} - 10^{13} h^{-1} M_\odot$
(solid, dashed and dotted lines respectively) for a $\Lambda$CDM
cosmology, with an initial adiabat $K_{34} = 0.3$ (in units of
$10^{34} $ erg cm$^2$ g$^{-5/3}$).  Right: The evolutionary tracks
along the $L$--$T$ relation for the halos on the left.  Large
squares and circles mark $z= 1$ and $0.5$ respectively.  Local data
from Allen \& Fabian (1998), Arnaud \& Evrard (1999), Ponman et
al. (1996).\label{fig1}}
\end{figure}

\section{The energy crisis}

The external heating scenario is appealing, but the
excess entropy cannot be spread uniformly into the cosmic baryons at
high redshifts, because this would make the Ly$_\alpha$ forest to
disappear.  Recently, it has been estimated that the level of
preheating in the Ly$_\alpha$ clouds is of the order of few $10^4 K$
(Cen \& Brian 2001), corresponding to an entropy level at least one
order of magnitude lower than that observed in groups.  A possible
solution is that only the baryons that end up in the cores of groups
and clusters are heated by a biased distribution of sources.  Such a
warm, low density gas would be unobservable at high $z$, and can be
detected as OVI absorption systems at low (Tripp, Savage \& Jenkins
2001) or at intermediate $z$ (Reimers et al. 2001).

\begin{figure}
\plottwo{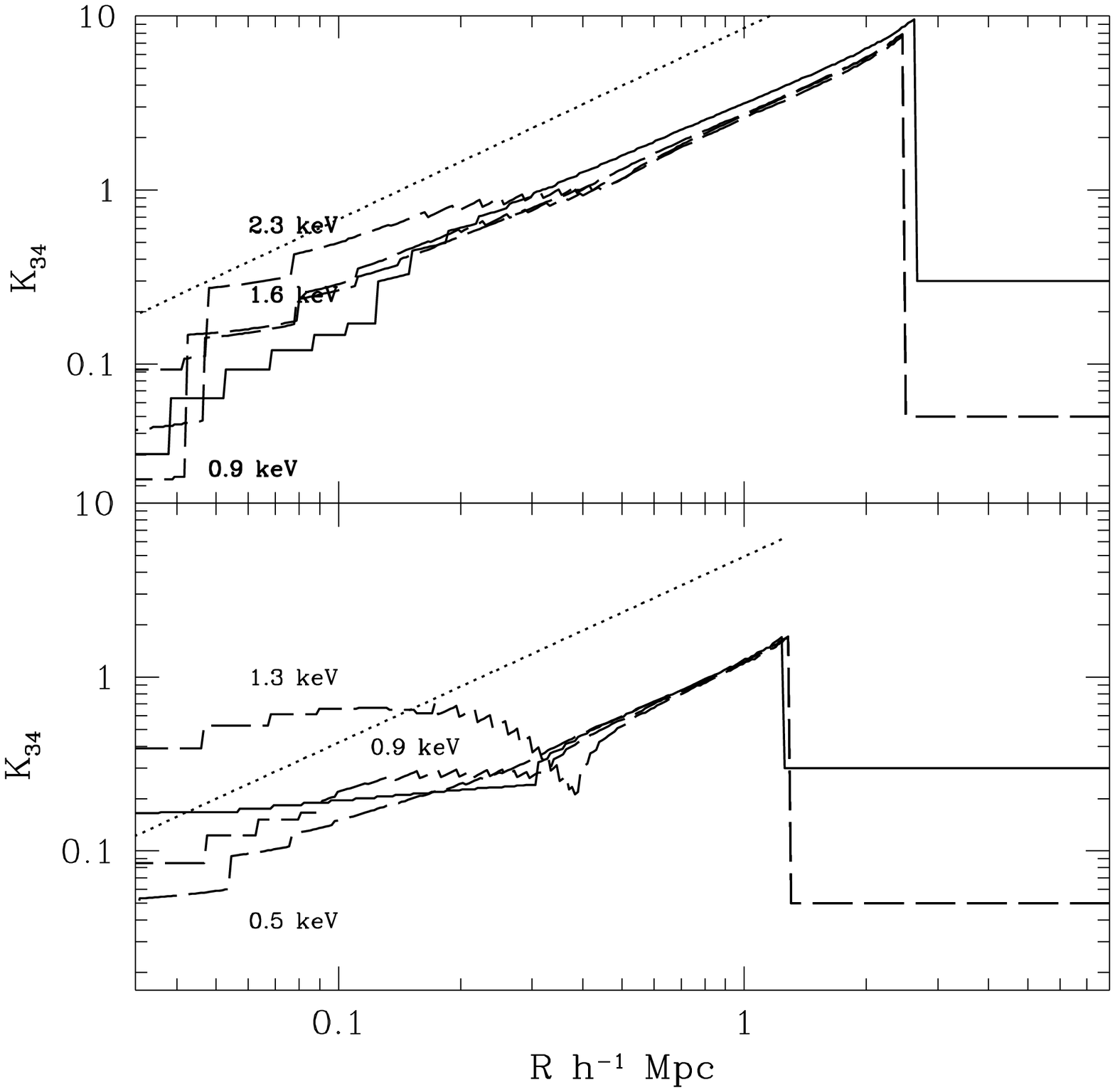}{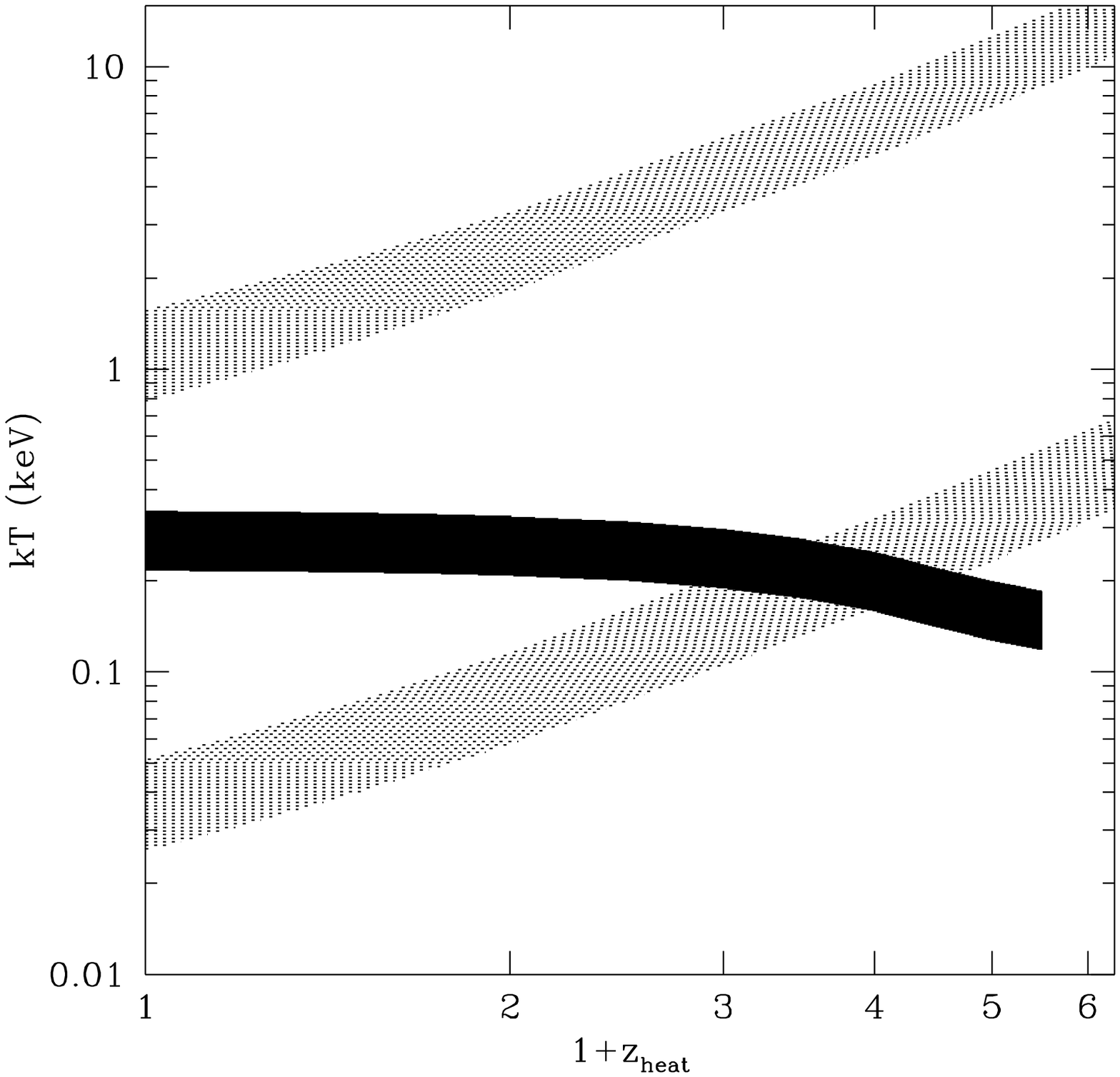}
\caption{Right: The entropy profiles in the external (solid) and the
internal (dashed lines) scenario for a rich ($M=10^{15} h^{-1}
M_\odot$, top) and a small cluster ($M=10^{14} h^{-1} M_\odot$,
bottom).  The external entropy level is $K_{34} = 0.3$.  In the
internal scenario, a total energy of 1--2 keV per particle is
released, as shown by the labels.  Left: The upper and lower stripes
show the required energy per particle needed to obtain an excess
entropy of $K_{34} = 0.2 - 0.3$ in virialized structures and in the
background baryons respectively.  The solid stripe is the energy per
particle dumped in the ICM by TypeII and TypeIa SNae after Pipino et
al. (2001).\label{fig2}}
\end{figure}

If OVI systems cannot account for the pre--heated baryons, the ICM
must be heated after the collapse.  Of course, for a given entropy
level, the much higher density of the collapsed regions implies a much
higher energy input.  An energy budget of 1--2 keV per particle (3--10
times higher than that in the external scenario) seems to be required
to reproduce the entropy floor (see also Wu, Fabian \& Nulsen 1999,
Valageas \& Silk 1999) and entropy profiles similar to the ones
predicted in the external heating scenario (see Figure 2, left).

Despite the SNae can provide a large amount of energy, their
efficiency in heating the gas is unknown.  In particular, if the
heated gas has high density, the thermal energy received from SNae is
rapidly radiated away, with a small net increase in the gas entropy.  A
recent study has been made by Pipino et al. (2001), starting from the
observed luminosity function of cluster galaxies.  The efficiency of
TypeII and TypeIa SNae is computed assuming a spherical gas
distribution around galaxies.  The energy per particle dumped in the
ICM as a function of redshift is plotted in Figure 2 (right, black
stripe).  It turns out that SNae can contribute a substantial amount
of the required energy ($\simeq 0.3$ keV/particle), which is,
however, lower than that needed in virialized region (upper stripe).
The first attempts to include the stellar feedback in hydrodynamical
simulations, seem to indicate a low efficiency from SNae (see, e.g.,
Borgani et al. 2001).  Following the natural predisposition of
cosmologists towards the use of the word {\sl crisis}, I would call
this empasse the {\sl energy crisis} of the cosmic baryons.

\begin{figure}
\plotone{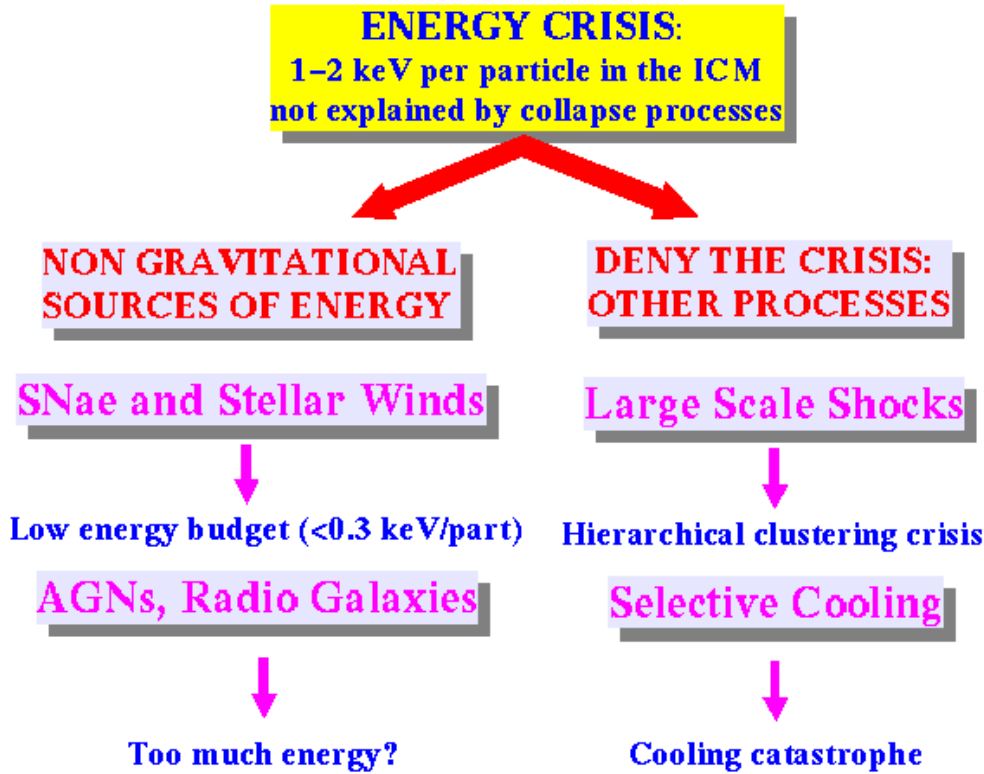}
\caption{A sketch of the current situation.  The description may be
not exhaustive.}
\label{fig3}
\end{figure}

\section{Solutions and possible way out}

If SNae will be shown unadequate for this job, the energy crisis can
be solved by other sources of non--gravitational energy.  Several
studies already considered AGNs as the main source of heating.  Radio
Galaxies can provide an energy output two orders of magnitude larger
than stellar sources, and they can heat the baryons at large distance,
reaching low densities for which the radiative cooling is negligible
(see Inoue \& Sasaki 2001).  On the other hand, AGNs can heat the
baryons so efficiently as to exceed the upper limit of the
comptonization parameter of the CMB (Yamada \& Fujita 2001).

As in political life, a possible way to solve the crisis is to deny
it: some viable scenario avoids the intervention of non--gravitational
heating to generate the entropy plateau.  The extra energy could be
due to gravitational shocks produced by large scale structure
(filaments and sheets) before the collapse of group--sized regions.
However, this requires large scale shocks to occur at redshifts much
larger than the typical formation epoch of the cores of clusters or
groups, which can be as high as $z\simeq 3$ in $\Lambda$CDM universes.
These large--scale effects have never been noticed in N--body
simulations with current CDM power spectra, for which there is a large
consensus.

Another way consists in simply eliminating the low entropy gas.  Some
simulations suggest that the cooling itself, mostly efficient in the
higher density region, can reproduce the X--ray properties of
clusters, like the $L$--$T$ relation (Muanwong et al. 2001).  In this
case, the entropy plateau is created by the removal of the lowest
entropy gas, which cools out of the diffuse, emitting phase.  However,
the majority of present--day simulations, in absence of any feedback,
find an unacceptably high fraction of cooled baryons (Balogh et
al. 2001).  We must stress that the cooling is extremely difficult to
treat numerically and there is consensus among simulations only at
very high resolutions.

To summarize, denying the crisis can lead to a worst crisis, like the
inversion of the clustering hierarchy or the cooling catastrophe, as
sketched in Figure 3.  At the present stage there are no obvious
solutions and it is important to investigate all the possible way out.
Both X--ray observations and theoretical modelling will be crucial in
the next years to get out of this empasse.


\begin{references}
\reference{} Allen, S.W., \& Fabian, A.C. 1998, MNRAS, 297, 63L
\reference{}Arnaud, M., \& Evrard, A.E. 1999, MNRAS, 305, 631
\reference{}Balogh, M., Pearce, F., Bower R., Kay S. 2001, MNRAS, 326,
1228
\reference{}Borgani, S. et al. 2001, ApJ in press, astro-ph/0108329 
\reference{} Cen, \& Bryan 2001, ApJ, 546, 81L 
\reference{} Inoue, S., \& Sasaki S. 2001, ApJ in press, astro-ph/0106187 
\reference{} Muanwong et al. 2001, ApJL, 552, L27
\reference{} Mushotzky, R.~F., \& Scharf, C.~A. 1997, ApJ, 482, 13
\reference{} Ponman, T.~J., et al. 1996, MNRAS, 283, 690 
\reference{} Ponman, T.~J., Cannon, D.~B., \& Navarro, F.J. 1999,
Nature, 397, 135 
\reference{} Reimers, D., et al. 2001, A\&A submitted, astro-ph/0106097
\reference{} Tozzi, P., \& Norman, C. 2001, ApJ, 546, 63 (TN)
\reference{} Tozzi, P., Scharf, C., \& Norman, C. 2000, ApJ, 542, 106 (TSN)
\reference{} Tripp, T.M., Savage, B.D., \& Jenkins, E.B. 2001, ApJL in press, astro-ph/0003277
\reference{}Valageas, P., \& Silk, J. 1999, A\&A, 350, 725
\reference{} Yamada, M., \& Fujita, Y. 2001, ApJL, 553, 14
\reference{}Wu, K.~K.~S., Fabian, A.~C., \& Nulsen, P.~E.~J. 1999, 
MNRAS, 318, 889

\end{references}
\end{document}